%% file: graphical-hc.tex
\documentclass[a4paper,reprint,showkeys,superscriptaddress,onecolumn,tightenlines,nofootinbib,11pt]{revtex4-2}

\usepackage{balance}
\usepackage{amsmath}
\usepackage{amssymb}
\usepackage{bbold}
\usepackage{graphicx}
\usepackage{subcaption}
\usepackage{mathtools}
\usepackage{float}

\usepackage{tikz}
\usetikzlibrary{decorations.markings,arrows.meta,math,calc}
\tikzset{centerorigin/.style={%
                              execute at end picture={%
                                                      \path let \p1=(current bounding box.west),\p2=(current bounding box.east)
                                                       in ({-max(-1*\x1,\x2)},\y1) ({max(-1*\x1,\x2)},\y1);
                                                     }
                              }
        }

\usepackage{chr}

\usepackage{etoolbox}
\makeatletter
\patchcmd{\frontmatter@abstract@produce}
  {\vskip200\p@\@plus1fil
   \penalty-200\relax
   \vskip-200\p@\@plus-1fil}
  {}
  {}
  {}
\makeatother

\hypersetup{
	pdfauthor={Faezeh Khodabandehlou, Simon Krekels and Irene Maes},
	pdftitle={A graphical approach to nonequilibrium heat capacities}
			}

\begin{document}
\title{Exact computation of heat capacities for active particles on a graph}
\author{Faezeh Khodabandehlou}\affiliation{Instituut voor Theoretische Fysica, KU Leuven}
\author{Simon Krekels}\email{simon.krekels@kuleuven.be}\affiliation{Instituut voor Theoretische Fysica, KU Leuven}
\author{Irene Maes}\affiliation{Departement Wiskunde, KU Leuven}

\date{\today}

\begin{abstract}
The notion of a nonequilibrium heat capacity is important for bio-energetics and for calorimetry of active materials more generally.   It centers around the notion of excess heat or excess work dissipated during a quasistatic relaxation between different nonequilibrium conditions.
We give exact results for active random walks moving in an energy landscape on a graph, based on calculations employing the matrix-tree and matrix-forest theorems.  
That graphical method applies to any Markov jump process under the physical condition of local detailed balance.
\end{abstract}
\maketitle

\section{Introduction}
From the conception of heat and thermodynamics, through the birth of quantum mechanics and all the way to black holes, heat capacities have played a central and foundational role in chemistry and physics \cite{einstein1907Plancksche,lynden-bell1999Negative,dinsmore2020Schottky}. 
Biology and complexity science have encountered fewer successes emerging from the study of heat capacities. 
On a fundamental level, such progress requires a theory of heat exchanges for nonequilibrium systems and for driven random motion in an energy landscape on complex networks.  
To that effect, we wish to understand quantitatively how changes in parameters such as environment, temperature, or activity modify the heat capacity in such systems.
Therefore, exact results concerning nonequilibrium heat capacities should be looked out for.  
Such is the motivation of this paper: to introduce nontrivial and interesting toy models of active systems on graphs and to give exact results for their heat capacity.

It is important to specify what is meant by a heat capacity for nonequilibrium systems.  
It is not a purely thermodynamic quantity and cannot, for example, be reduced to the temperature-derivative at fixed volume of the mean energy.
Rather, what is needed is a proper extension of the equilibrium notion of heat capacity, which makes use of the concept of excess heat.   
Such an extension was pioneered in \cite{boksenbojm2011Heat,pesek2012Model,maes2019Nonequilibrium}, building on earlier ideas by e.g.~\cite{oono1998Steady,glansdorff1954Proprietes,maes2015Revisiting,komatsu2008SteadyState}.   
In the next section we briefly sketch the main ingredients that go into the definition of excess heat and heat capacity.  
We then apply it to the case of an active random walker on a graph imbued with an energy landscape.

Central in this paper is a graphical method based on matrix-tree and matrix-forest theorems, extensions in a way of the well-known Kirchhoff formula for the stationary distribution of a Markov process.
The graphical method relies on the enumeration of trees and forests, which generally makes the algorithm complexity increase exponentially with the graph size, quickly becoming unfeasible (more than $\sim\!20$ vertices).
That method and its algorithmic implementation are discussed in Section \ref{sec:alg}. It is worth emphasizing that the methods described below are applicable to a much broader category of systems: we can also imagine the graph to represent lattices or networks for transmission of matter or information.  
Even more generally, the graphs may represent any Markov jump process with a discrete configuration space \cite{Ir}.
However, we restrict our scope and focus on exact computations pertaining to active systems.
Even though collective phenomena such as phase transitions remain beyond our scope, small active systems are already interesting for their behavior of the heat capacity as function of temperature, persistence, energy and activity parameters. 
Section \ref{sec:res} presents several examples and we illuminate the relevant features in their heat capacities, which are calculated exactly for all temperatures. 
These results may be compared to those from simulation methods for obtaining the heat capacity of thermal run-and-tumble particles diffusing in one-dimensional energy landscapes are reported in \cite{dolai2022active}.

\section{Excess work and heat capacity}

\subsection{General idea}

\begin{figure}
    \centering
    \includegraphics[width=0.5\linewidth]{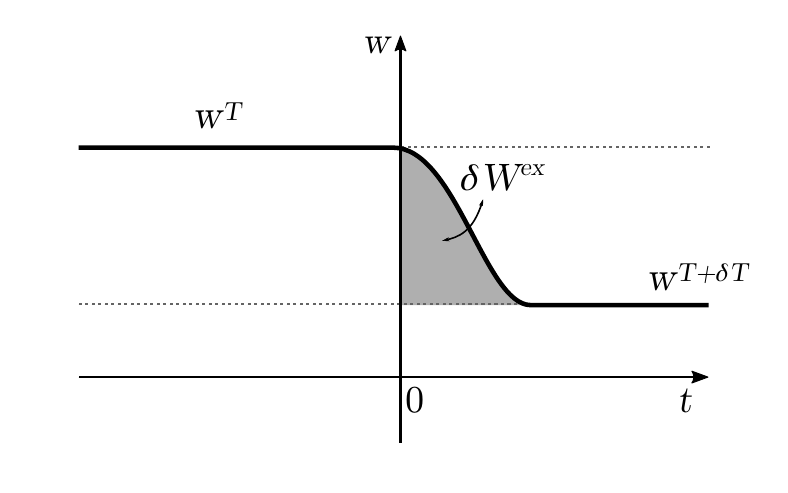}
    \caption{A schematic plot of the work as function of time, for a sudden temperature change at $t=0$. The shaded area represents the excess work produced with a change in bath temperature $\delta T$ at time zero.  The
    $w^T$ and $w^{T+\delta T}$ are the stationary work performed before $t=0$ and in the far future.}
    \label{fig:Qex}
\end{figure}

We refer to \cite{boksenbojm2011Heat,pesek2012Model,maes2019Nonequilibrium} for details concerning the theory of nonequilibrium heat capacities.
The main idea is to identify the heat capacity $C(T)$ with the excess heat $\delta Q^{\mathrm{ex}}$ exchanged with the system for accommodating a (small) change $\delta T$ at environment temperature $T$. 

Indeed, imagine an open system in contact with a thermal bath in equilibrium of which the temperature $T$ can be controlled.
To keep the system in a steady nonequilibrium state,  external forces continuously exert power with average $w^T$ on the system by, for example, pushing particles in a given direction. 
Now assume that the system undergoes such a sudden and small change in temperature, say from  $T$ to $T+\delta T$ at time $t=0$.
Then, the system relaxes to the new steady condition with bath temperature $T+\delta T$ and with the new dissipated power $w^{T+\delta T}$.
During the relaxation and in the interval $[0,t]$, the work done on the system is $W_{[0,t]}$, always obeying with the First Law
\begin{equation}\label{eq:1l}
W_{[0,t]}+Q_{[0,t]} = E(t)-E(0)
\end{equation}
where $E$ is the energy of the system and $Q_{[0,t]}$ is the heat exchanged with the system.   
There is an \emph{excess work}, entirely due to change in temperature which equals 
\begin{equation}
    \delta W^{\mathrm{ex}} = \lim_{t\to\infty}\left(W_{[0,t]}-w^{T+\delta T}t\right)
\end{equation}
see Fig.~\ref{fig:Qex}.
In terms of that excess work, the steady heat capacity is
\begin{equation}\label{eq:hc_wex}
    C(T) = \frac{\partial E}{\partial T} - \frac{\delta W^{\mathrm{ex}}}{\delta T}.
\end{equation}
Indeed $C(T)\dif T = \delta Q^{\mathrm{ex}}$ as can be seen by invoking the First Law \eqref{eq:1l}.
The first term in the right-hand side of \eqref{eq:hc_wex} takes the form of the usual equilibrium heat capacity at fixed volume; in equilibrium the second term vanishes.

It is worth mentioning that the excess heat is not unique, see for example \cite{PhysRevLett.86.3463}. Note, however, that we consider an excess heat which is geometric and reversible. 
These are nontrivial properties and exactly fit what could be expected from equilibrium thermodynamics. 
Another reason for our choice of excess heat, is that it avoids discussions related to entropy and the second law. For further discussion on this issue, see  \cite{let, mathnernst}. 

\subsection{Active random walker on a graph}
We now restrict ourselves to systems which we call \emph{active systems} on graphs.
Consider a finite, connected graph $G$, which has vertices $x,y,\ldots$ and directed edges $(x,y)$ from $x$ to $y$.
The vertices have an energy $E(x)$ associated to them, and each directed edge has a rate $k(x,y)$ associated with it.  

Let $X_t$ be a random walker in an energy landscape on the vertices of the graph $G$.
In addition, there is a force acting on $X_t$, which depends on an \emph{internal state} $\sigma$.
Specifically, this internal state is represented by a dichotomous noise, $\sigma \in \{+1,-1\}$, occasionally flipping from $+1$ to $-1$ according to a (fixed and constant) rate $\alpha$.
The value of $\sigma$ affects the transition rates, which we acknowledge by adding a subscript: $k_\sigma(x,y)$.
Crucially, the compound process $(X_t, \sigma_t)$ for the dynamics will be Markovian.
We thus speak of an \emph{active} random walker.
Graphically, we imagine two duplicate graphs, $G_+$ and $G_-$, with identical structure but different transition rates according to whether $\sigma=\pm1$.
In particular, an ordinary random walker on $G_+ \cup G_-$, where $G_\pm$ are connected by edges with rate $\alpha$, represents the active random walker on $G$.

The random walker is in contact with a thermal bath in equilibrium at inverse temperature $\beta=\frac{1}{T}$, letting $k_B=1$. 
The active force performs work $W_\sigma(x,y)=-W_\sigma(y,x)$ on the walker during the transition $x\to y$, which is instantly dissipated into the thermal bath.
Effectively, we assume that the system obeys \emph{local detailed balance} \cite{maes2021Local}, requiring
\begin{equation}\label{eq:local_db}
    \log \frac{k_\sigma(x,y)}{k_\sigma(y,x)} = \beta\,[E(x)-E(y) + W_{\sigma}(x,y)].
\end{equation}
Section \ref{sec:res} provides explicit forms of $k_\sigma(x,y)$ in terms of $E(x)$, $E(y)$ and $W_\sigma(x,y)$.

In the transition $x\to y$, the heat dissipated to the reference bath  is
\begin{equation}\label{fig:1st_law}
    Q_\sigma(x,y) = E(x)-E(y) + W_{\sigma}(x,y)
\end{equation}
which is a differential form of the First Law \eqref{eq:1l} at fixed $\sigma$.
As for many cases for active systems with dichotomous noise, we let
\begin{equation}\label{eq:dich_work}
    W_{\sigma}(x,y) = \sigma\, \mathcal{E}(x,y),\quad \mathcal{E}(x,y) = - \mathcal{E}(y,x)
\end{equation}
meaning that $W_{+}(x,y) + W_{-}(x,y) =0$, thus eliminating any net global bias.

The expected power exerted on the system when in state $x$ is
\begin{equation}\label{eq:expected_power}
w(x,\sigma) = \sum_y k_\sigma(x,y) W_{\sigma}(x,y)\,.
\end{equation}
Now let the walker $X_t$ start at position $x$  and in (internal) state $\sigma$ at $t=0$.
At bath temperature $T+\dif T$, the expected power at time $t$ is
\begin{equation}
\big\langle w(X_t,\sigma_t)\,|\,(X_0,\sigma_0)=(x,\sigma)\big\rangle_{T+\dif T}\,.
\end{equation}
The excess work is
\begin{equation}\label{eq:ex_work_int}
V^{T+\dif T}(x,\sigma) = \int_0^\infty \left[\big\langle w(X_t,\sigma_t)\,|\,(X_0,\sigma_0)=(x,\sigma)\big\rangle_{T+\dif T} - w^{T+\dif T}\right]\dif t
\end{equation}
where we subtracted the asymptotic stationary value
\begin{equation}
\big\langle w(X_t,\sigma_t)\,|\,(X_0,\sigma_0)=(x,\sigma)\big\rangle_{T+\dif T}\xrightarrow[]{t\uparrow \infty} w^{T+\dif T}.
\end{equation}
The convergence is exponential which makes the integral \eqref{eq:ex_work_int} well-defined.  
Note that $V^{T+\dif T}$ depends on the final temperature $T+\dif T$.  On the other hand $\langle V^{T}\rangle_{T} =0$. Therefore, following \eqref{eq:hc_wex}, 
 the heat capacity at temperature $T$ is
\begin{equation}\label{eq:heatcapacity}
    \begin{aligned}
        C(T) &= \frac{1}{\dif T}[ \langle E\rangle_{T+\dif T} -\langle E\rangle_T - \langle V^{T+\dif T}\rangle_T \,]\\
             &= \frac{\dif\langle E\rangle_T}{\dif T} - \Big\langle\frac{\dif V^{T}}{\dif T}\Big\rangle_T
    \end{aligned}
\end{equation}
where $\langle E\rangle_T$ is the average energy.

\section{Graphical representation}\label{sec:alg}

The introduction of the active noise $\sigma$ leads to a \emph{double graph} structure, where each vertex is mirrored by another, but the edges between them have rates where $\sigma$ is inverted.
The rates of flipping between the graphs is $\alpha$.
Consider the simple example of a two-level system, which is represented by a graph with two connected vertices.
Adding dichotomous noise will ``double'' the graph to one with four vertices, connected by edges with rate $\alpha$.
\begin{center}
    \input{figures/double_graph}
\end{center}
The active process $\mathcal{I}_t=(X_t,\sigma_t)$ is thus a Markov jump process on a finite simple and connected graph with vertices (states) $i$ labeled as $i =(x,\sigma)$.
The transition rates $\lambda$ for the  Markov process $(X_t,\sigma_t)$ on the double graph are
\begin{equation*}
    \lambda(i,j) = \begin{cases}
    k_\sigma(x,y) & \text{for}\ i=(x,\sigma),\ j=(y,\sigma)\\
    \alpha & \text{for}\ i=(x,\sigma),\ j=(x,-\sigma)
    \end{cases}
\end{equation*}
and the backward generator has matrix elements
\begin{equation}
    \begin{aligned}
        L_{ij} &= \lambda(i, j)\\
        L_{ii} &= -\sum_{j \neq i} \lambda(i, j).
    \end{aligned}\label{L}
\end{equation}
We encapsulate the expected power \eqref{eq:expected_power} by considering the double graph
\begin{equation}
    w(i) = \sum_j \lambda(i,j)W(i,j).
\end{equation}
with 
\begin{equation}
    W(i,j) = \begin{cases}
    \sigma \mathcal{E}(x,y) & \text{for}\ i=(x,\sigma),\ j=(y,\sigma)\\
    0 & \text{for}\ i=(x,\sigma),\ j=(x,-\sigma).
    \end{cases}
\end{equation}
We assume here that tumbling ($\sigma\to-\sigma$) produces no heat in the thermal environment and, as in \eqref{eq:dich_work}, $\mathcal{E}(x,y)=-\mathcal{E}(y,x)$.\\

To calculate the heat capacity from \eqref{eq:heatcapacity}, we need the stationary distribution $\rho^T(x,\sigma)$ and the excess work $V^T(x,\sigma)$.
To this end, we employ \emph{graphical} methods, the first of which is commonly called the  matrix-tree theorem.
The second method makes use of the matrix-forest theorem.

\subsection{Stationary distribution}

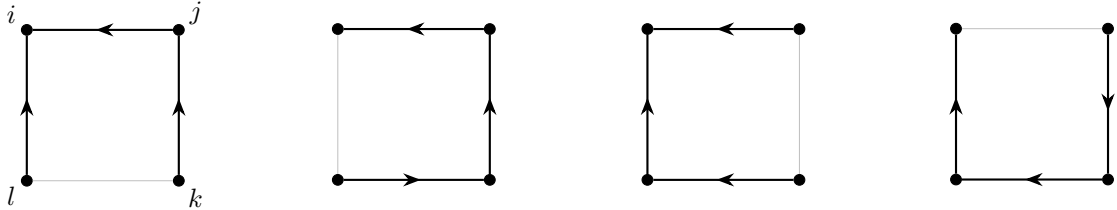
\begin{figure*}
    \centering
    \input{figures/spanning_trees_4}
    \caption{All spanning trees rooted in $i$ for a graph representing an $n=4$ Markov jump process.}
    \label{fig:spanning_trees}
\end{figure*} 

The stationary distribution $\rho(x,\sigma) =\rho(i)$ is the normalized row vector that solves the stationary master equation
\begin{equation}\label{eq:stationary_master}
    \rho\, L = 0.
\end{equation}
The matrix-tree theorem states that the solution to \eqref{eq:stationary_master} may be expressed in terms of weights of the spanning trees of that graph \cite{shubert1975FlowGraph, anantharam1989proof,tutte1948dissection}.
The Kirchhoff formula, making use of the matrix-tree theorem, gives a way to represent the solution to equation \eqref{eq:stationary_master} in terms of spanning tree weights
\begin{equation}\label{eq:graph_rep_rho}
    \rho(i) = \frac{\mathcal{M}(\mathcal{T}_i)}{\sum_i \mathcal{M}(\mathcal{T}_i)}.
\end{equation}
By a tree, we mean a connected cycle-less component of a graph, and a spanning tree is a tree including all vertices of the graph. 
A tree can be defined by giving all of its constituent edges.

The weight $m$ of a spanning tree $\mathcal T$ is simply the product of the rates associated to all edges in that tree:
\begin{align}
    m(\mathcal T) = \prod_{(u, u') \in \mathcal T} \lambda(u, u').
\end{align}

Let the set of all rooted spanning trees rooted in $i$ be denoted $\mathcal{T}_i$.
Here, \emph{rooted} means that all edges flow to the same point, the root.
The weight $\mathcal{M}$ of such a set of trees is the sum of the weights of the trees:
\begin{equation}
    \begin{aligned}
        \mathcal{M}(\mathcal{T}_i) &= \sum_{\mathcal T\in\mathcal{T}_i}m(\mathcal T) \\
        &= \sum_{\mathcal T\in\mathcal{T}_i}\prod_{(u, u')\in \mathcal T}\lambda(u, u').
    \end{aligned}
\end{equation}

Thus, to find $\rho(i)$, we simply need to find the sets of all spanning trees rooted in $i$ for all $i$; a simple example is shown in Fig.~\ref{fig:spanning_trees},  for more examples and details see \cite{intro}.
These trees can be generated algorithmically, yet the computational complexity increases sharply with the number of states.

\subsection{Excess work}

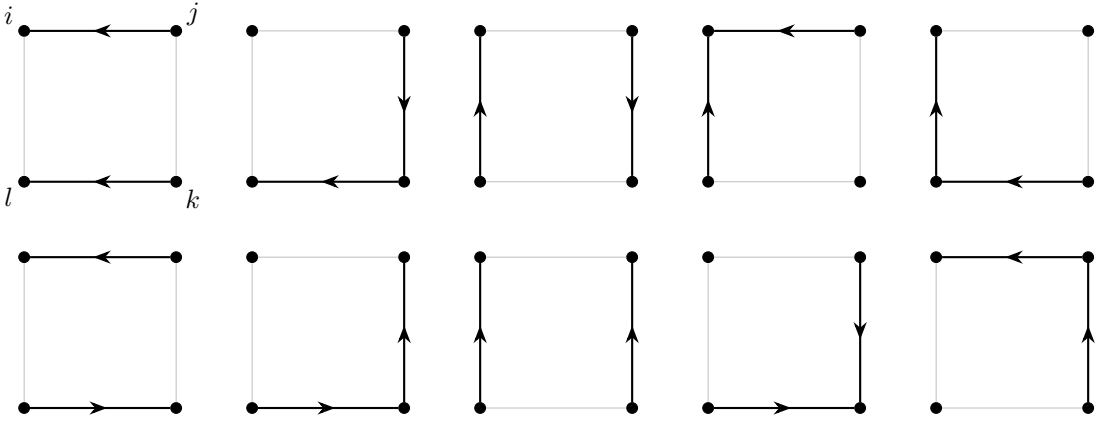
\begin{figure*}
    \centering
    \input{figures/spanning_forests_4}
    \caption{The spanning forests $\mathcal{F}^{i\to i}$ for a graph representing an $n=4$ Markov jump process. The sets $\mathcal{F}^{u\to i}$ for $u\neq i$ are all subsets of $\mathcal{F}^{i\to i}$.}
    \label{fig:spanning_forests}
\end{figure*}

The excess work \eqref{eq:ex_work_int} can be graphically expressed using the \emph{Matrix-Forest} theorem \cite{agaev2006Matrix}.
Omitting sub- and superscripts now to indicate temperatures, we use that
\begin{equation}
    \langle w(\mathcal{I}_t)\,|\,\mathcal{I}_0=i\rangle = e^{Lt}w(i).
\end{equation}
The excess work \eqref{eq:ex_work_int} can then be written as
\begin{align}\nonumber
    V(i) &= \int_0^{\infty} [e^{Lt}w(i) - \left< w \right>] \dif t\\\nonumber
    & =  \int_0^{\infty}  e^{Lt} [w(i) - \left< w \right>]\dif t\\\label{eq:exc_w_exp}
    & = \int_0^{\infty} e^{Lt} f(i)\,\dif t
\end{align}
where $f(i) = w(i) - \left< w \right>$ is the zero-average expected work.

The last expression \eqref{eq:exc_w_exp} is also equal to the limit of the resolvent inverse acting on $f$ \cite{taylor2011Partial}
\begin{equation}
   V(i) = \int_0^{\infty} e^{Lt} f(i)\,\dif t = \left(\lim_{b \to \infty} b (I - b L )^{-1} f\right)\!(i)\,.
\end{equation}
The matrix-forest theorem \cite{chebotarev2006Matrices,chebotarev2006MatrixForest} gives a graphical representation of $(I - b L)^{-1}$ in terms of $b$ and weights of so-called \emph{forests}. 
Using this representation 
\begin{align*}
   V(i) & =\sum_{j} \left(\lim_{b\rightarrow \infty} \dfrac{b}{ I-b L}\right)_{ij} f(j) \\
        & =\sum_{j} \lim_{b \to \infty} \frac{\sum_{k=0}^{n-1} b^{k+1}\mathcal{M} (\mathcal{F}^{i \to j}_k) f(j)}{\sum_{k=0}^{n-1}b^k \mathcal{M} (\mathcal{F}_k)}\\
        & =\sum_{j} \lim_{b \to \infty} \frac{\sum_{k=0}^{n-2} b^{k+1}\mathcal{M} (\mathcal{F}^{i\to j}_k) f(j)}{\sum_{k=0}^{n-1}b^k \mathcal{M}(\mathcal{F}_k)} + \sum_{j} \lim_{b \to \infty} \frac{b^{n}\mathcal{M}(\mathcal{F}^{i \to y}_{n-1})f(j)}{\sum_{k=0}^{n-1}b^k\mathcal{M}(\mathcal{F}_k)}
\end{align*}
$f$ has zero average, so the second sum vanishes.  
More details on the use of the matrix-forest theorem can be found in \cite{mathnernst}.
The result is an expression for $V$ in terms of the weights of forests :
\begin{equation} \label{eq:graph_rep_V}
    V(i) = \frac{\sum_{j} \mathcal{M}(\mathcal{F}^{i\to j}) f(j)}{\sum_j \mathcal{M}(\mathcal{T}_j)}.
\end{equation}
The forests and weights of forests are defined as follows: a forest is a cycle-less graph, whose connected components are trees; a \emph{spanning} forest is a forest which includes all vertices in the graph.
A forest is a \emph{rooted} forest if all the constituent trees are rooted.
The weight $m$ of a forest $F$ is the product of the rates of all the edges included in the forest,
\begin{equation}
    m(F) = \prod_{(u, u') \in F}\lambda(u,u').
\end{equation}
Let $\mathcal{F}_k^{i\to j}$ denote the set of all spanning forests consisting of $k$ edges such that $j$ is the root of one of the trees and $i$ belongs to the same tree as $j$. When $k=n-2$, every forest in $\mathcal{F}_{n-2}^{i\to j}$ consists of two trees and we use the notation $\mathcal{F}^{i\to j}$ instead of $\mathcal{F}_{n-2}^{i\to j}$. When $k=n-1$, every forest in $\mathcal{F}_{n-1}^{i\to j}$ consists of one tree which is a spanning tree. 
The weight $\mathcal{M}$ of a set of forests is the sum of the weights of the forests constituting the set,
\begin{equation}
    \begin{aligned}
        \mathcal{M}(\mathcal{F}^{i\to j}) &= \sum_{F\in\mathcal{F}^{i\to j}}m(F) \\
        &= \sum_{F\in\mathcal{F}^{i\to j}}\prod_{(u, u')\in F}\lambda(u, u').
    \end{aligned}
\end{equation}
Thus, to calculate the excess work one needs knowledge of the spanning trees of the graph and also the spanning forests of the graph.
As a simple example, see Fig.~\ref{fig:spanning_forests}; note that the lone vertices are to be seen as rooted trees, so there are always two trees in every forest.

\section{Examples and results}\label{sec:res}
So far we have not specified the transition rates $k_\sigma(x,y)$ except by the requirement of local detailed balance \eqref{eq:local_db}.
We choose rates of the general form
\begin{equation}\label{eq:bounded_rate_general}
    k_\sigma(x,y) = \frac{1}{ 1 + e^{-\beta [E(x) - E(y) + W_\sigma(x,y)]}}\,
\end{equation}
compatible with Arrhenius-type reaction rates \cite{aquilanti2017Kinetics}, and depending on the energy landscape $E(x)$ and the active work $W_\sigma(x,y)=\sigma\mathcal{E}(x,y)=-\sigma\mathcal{E}(y,x)$.  Other choices of rates are possible, and in contrast with equilibrium, the heat capacity may very well depend on the exact form of the transition rates even though the same local detailed balance condition is imposed.

\subsection{Active motion on a ring}
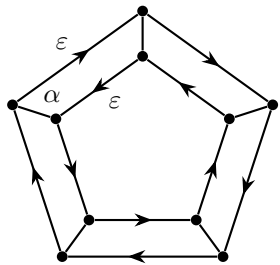
\begin{figure}[H]
    \centering
    \input{figures/dring}
    \caption{Double ring with $N = 5$, making the total number of vertices $n = 10$.}
    \label{fig:dring_graph}
\end{figure}

Consider a run-and-tumble particle on a periodic line (i.e., a ring) with $N$ sites.
Such a process can be represented by doubling the ring of $N$-sites to describe two different directions of driving.
These two rings are connected by edges representing the tumble transitions which occur at a fixed rate $\alpha$; the connected states are naturally at the same energy.
Thus in total there will be $n = 2N$ possible states, corresponding to the particle having clockwise or counterclockwise active orientation, see Fig.~\ref{fig:dring_graph}.

The transition rates are given by \eqref{eq:bounded_rate_general} for the edges drawn with arrows in Fig.~\ref{fig:dring_graph}, where $W_\sigma(x,y)=\pm\sigma\varepsilon$ depending on whether the edge is directed along or against the drawn arrow respectively.
We plot the heat capacity of the active walker, represented by the double ring with $N=5$ for different $\alpha$ and $\varepsilon$.
The energy function in each ring is 
\begin{equation}\label{eq:dring_energy}
    E(x)= E_0 \, \sin\left(\frac{2 \pi}{N}\,x\right)
\end{equation}
where $x\in\{0,1,\ldots, N-1\}$. 
\begin{figure*}[]
    \centering
    \begin{subfigure}{.49\linewidth}
    \centering
    \includegraphics[width=\linewidth]{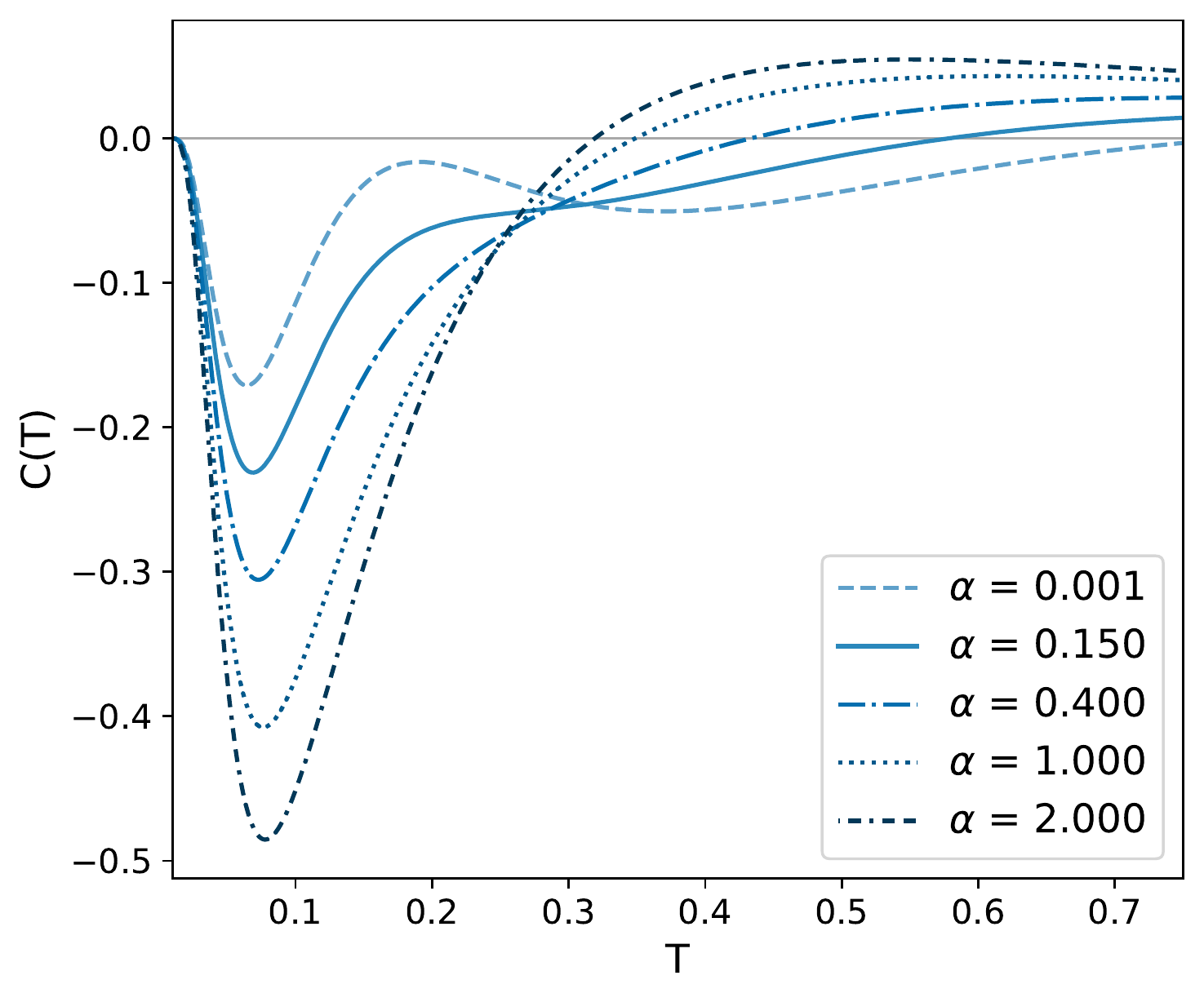}
    \caption{Varying $\alpha$,   $\varepsilon=1$.}
    \end{subfigure}
    \hfill
    \begin{subfigure}{.49\linewidth}
    \centering
    \includegraphics[width=\linewidth]{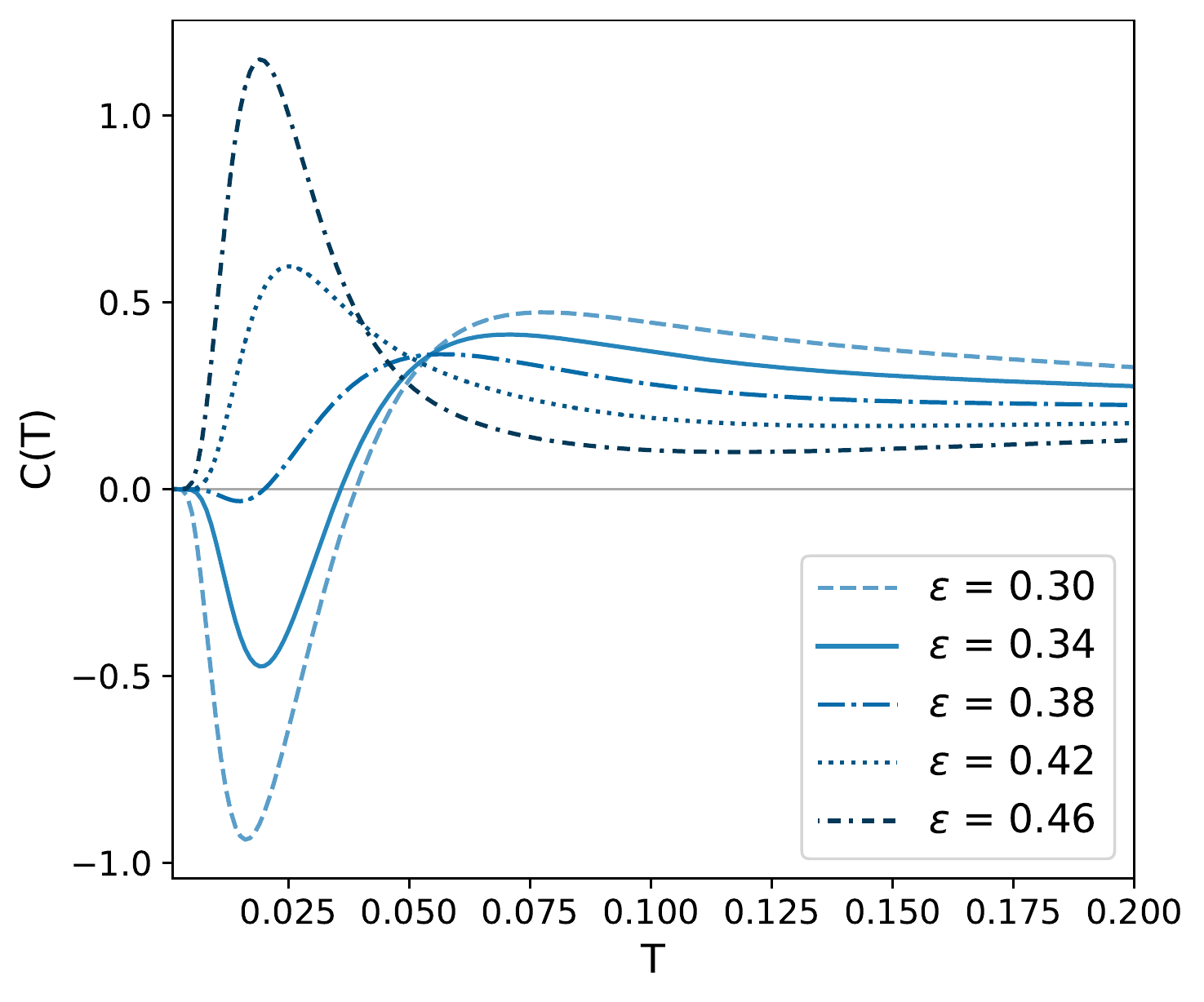}
    \caption{Varying $\varepsilon$,   $\alpha=0.5$.}
    \label{fig:dring_N5_eps}
    \end{subfigure}
    \caption{Heat capacities plotted for a range of temperatures for an active particle on an $N=5$ ring; $E_0=0.3$ in \eqref{eq:dring_energy}.}
    \label{fig:dring_N5}
\end{figure*}

\begin{figure}
    \centering
    \includegraphics[width=.5\linewidth]{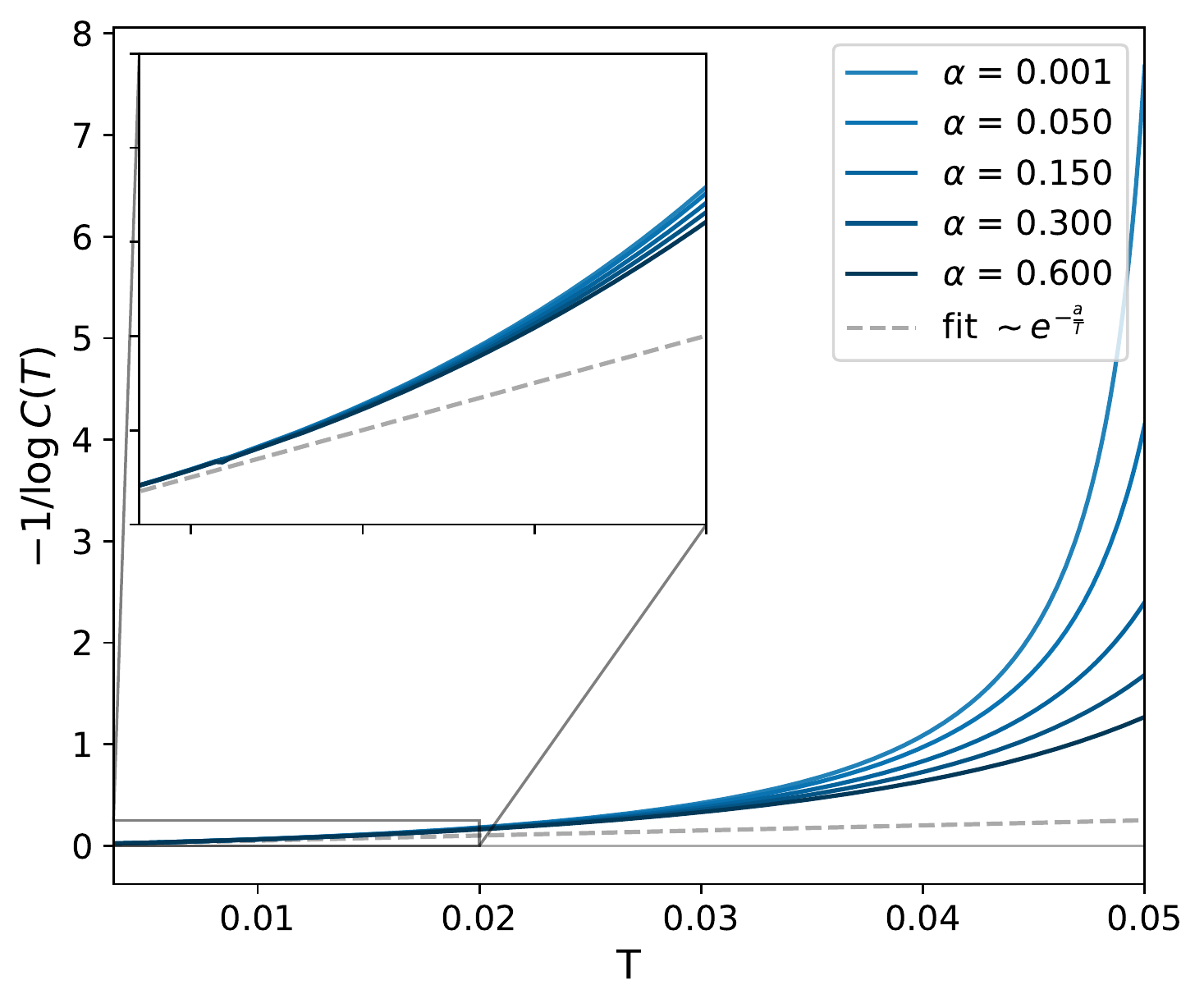}
    \caption{$C(T)$ approaches zero exponentially as $T\to0$. For low temperature, $C(T)$ decreases as $e^{-a/T}$ where $a\approx0.2$ is a fitted parameter. This plot shows $-1/\log\,C(T)$ for the $N=5$ multi-level system, which becomes approximately linear for small $T$.}
    \label{fig:exp_approach}
\end{figure}

In Fig.~\ref{fig:dring_N5} the results for $C(T)$ are shown. 
When the temperature approaches zero, the heat capacity also goes to zero exponentially \cite{mathnernst}, see for example Fig.~\ref{fig:exp_approach}. 
At low temperatures, the heat capacity can be negative and always reaches a minimum. 
In equilibrium, peaks in the heat capacity are referred to as Schottky anomalies; see e.g.~\cite{souza2016Specific}.
The peak is largest in absolute value for the largest values of $\alpha$, but saturates around $C(T^*)\approx-0.6$ for $\alpha\gtrsim2$. 
The temperature at which the minimum occurs is also $\alpha$-dependent, saturating around $T\approx0.08$. 
At higher temperatures, all heat capacities again become positive, and are larger for larger values of $\alpha$.

Varying $\varepsilon$  (Fig.~\ref{fig:dring_N5_eps}) gives widely different qualitative results, even for small changes in $\varepsilon$.
That is because in the rate function \eqref{eq:bounded_rate_general}, the driving strength $\varepsilon$ is divided by the temperature. 
Thus, as $T$ becomes small, small changes in $\varepsilon$ make large differences in the transition rates.
\subsection{Sisyphus random walk}

\begin{figure*}
    \centering
    \begin{subfigure}{.49\linewidth}
    \centering
    \includegraphics[width=\linewidth]{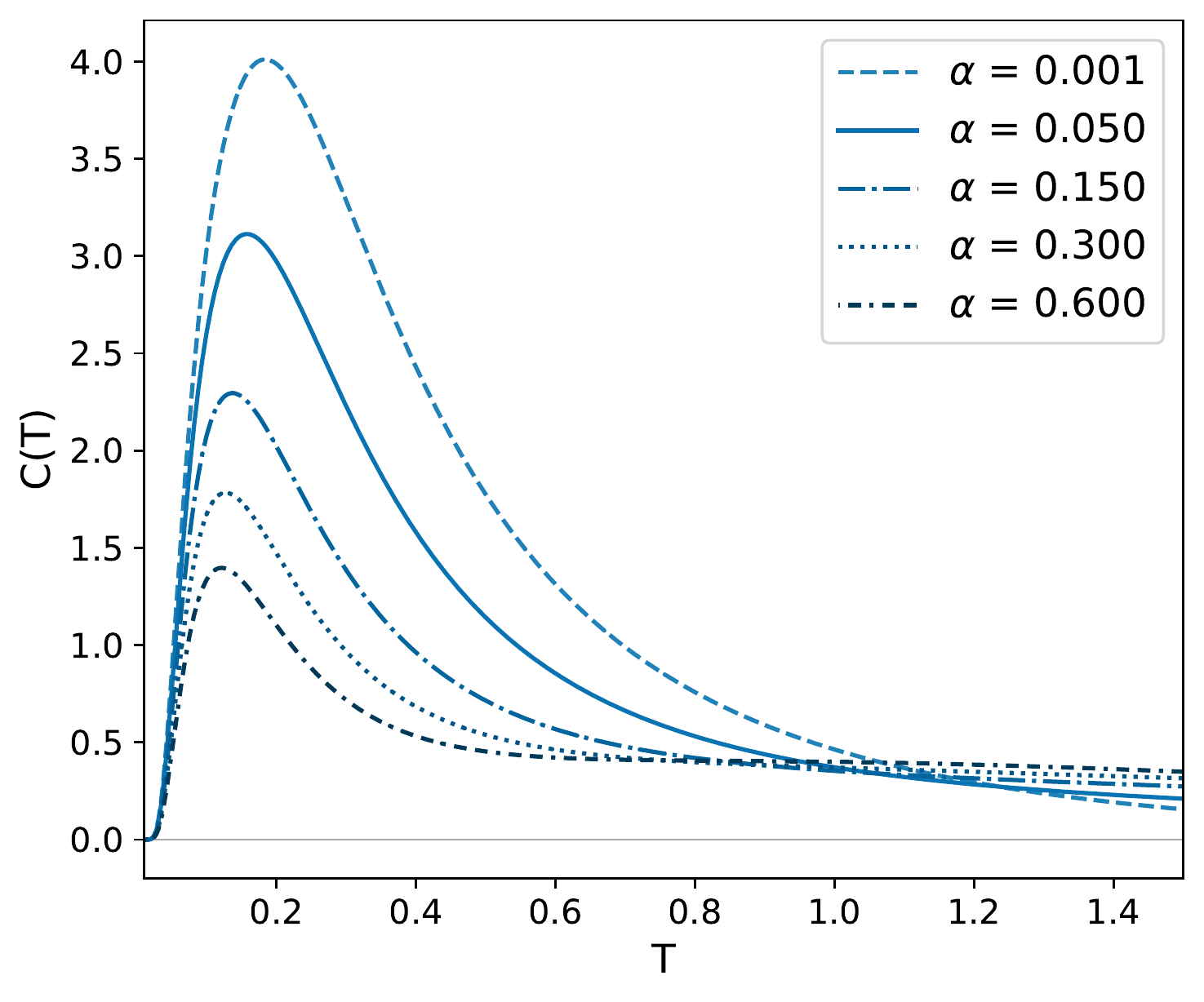}
    \caption{Varying $\alpha$,   $\varepsilon=1$.}
    \label{fig:ladder_alpha}
    \end{subfigure}
    \begin{subfigure}{.49\linewidth}
    \centering
    \includegraphics[width=\linewidth]{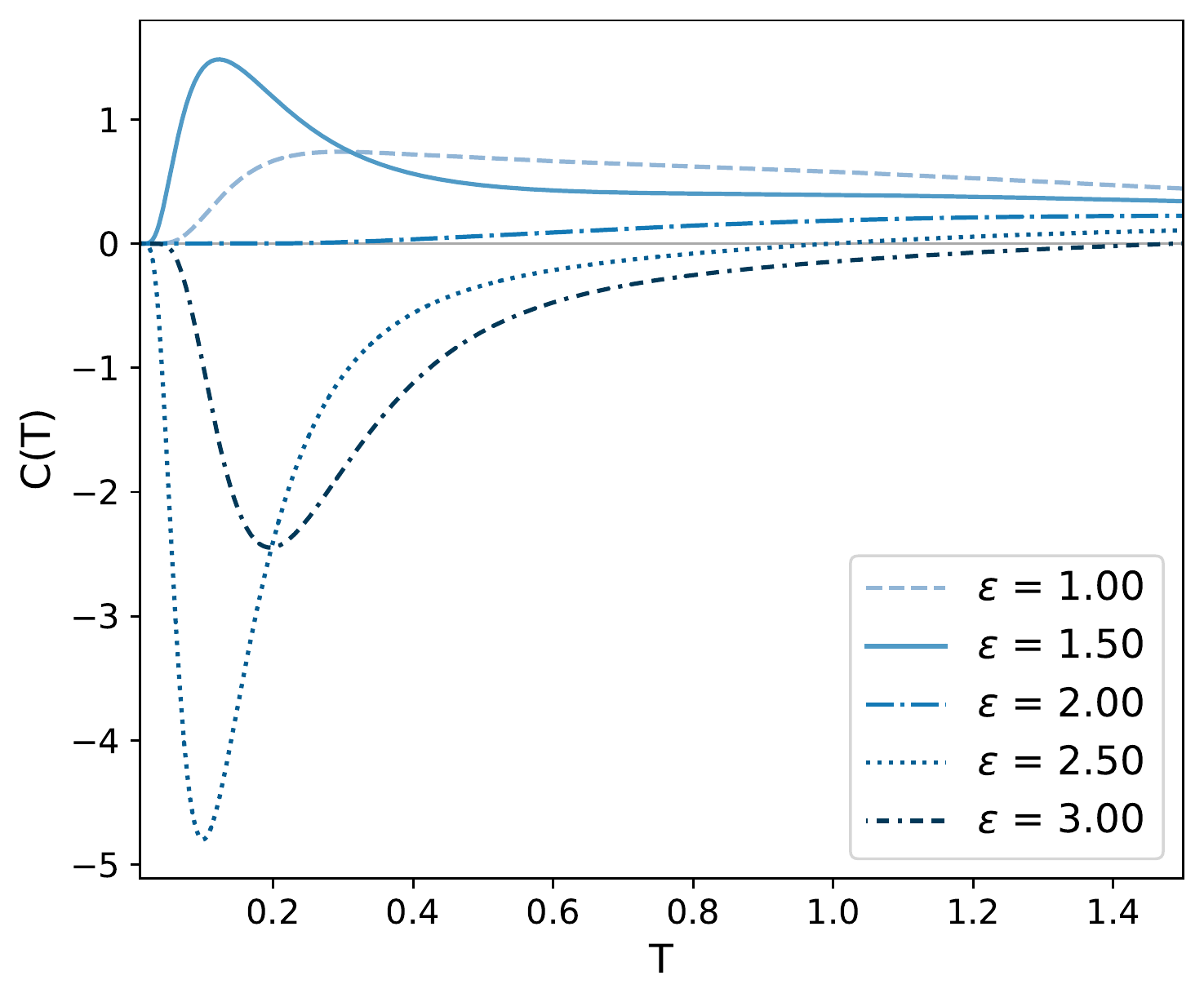}
    \caption{Varying $\varepsilon$,   $\alpha=0.5$.}
    \label{fig:ladder_eps}
    \end{subfigure}
    \caption{Heat capacity $C(T)$ of active walker on a  five-level ladder $(n=10)$ as function of temperature $T$.}
    \label{fig:ladder}
\end{figure*}

\begin{figure}[H]
    \centering
    \input{figures/ladder}
    \captionof{figure}{Energy-ladder}
    \label{fig:ladder_graph}
\end{figure}
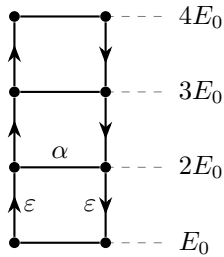

Consider a multi-level system with a switch.  
The switch reverses the direction of the driving.  
The corresponding graph representation is shown for $N = 4$ in Fig.~\ref{fig:ladder}.  
The situation is similar to the previous section, but now with an active walker on a line segment, with the energy increasing monotonically with $x$.
If we imagine the (double) line segment to be directed upwards, such as in Fig.~\ref{fig:ladder_graph}, the active walker is reminiscent of Sisyphus pushing his boulder up the mountain, only to tumble down again at a rate $\alpha$ \cite{montero2016Directed}.

The rates are again determined by \eqref{eq:bounded_rate_general} for the edges with arrows in Fig.~\ref{fig:ladder_graph}, where again $W_\sigma(x,y)=\pm\sigma\varepsilon$ depending on whether the walker moves along or against the arrows.

The graph representation of this Markov process is a ladder (Fig.~\ref{fig:ladder_graph}), i.e. two lines, one with an upward bias and one with a downward bias.
The two lines are connected by the ``tumbling'' transitions, where the bias of the walker is inverted at a rate $\alpha$. 
We plot the heat capacity for a five-level-ladder ($n=10$) when the energy spacings are equal and  $E_0=1$, the result is shown in Fig.~\ref{fig:ladder}. 

In  Fig.~\ref{fig:ladder_alpha}, we plot the heat capacity for different $\alpha$ and fixed $\varepsilon = 1$. 
The heat capacity is positive for the different values of $\alpha$ and it vanishes for temperature going to zero. 
For every $\alpha$, there is a maximum appearing in the low-temperature region. 
Increasing $\alpha$, we see that the peak moves to lower temperatures. 
At low temperatures, the heat capacity is largest for the smallest $\alpha$ and at high temperature it is the other way round.
For high $\alpha$ the heat capacity has two local maxima, the second being less pronounced.

In Fig.~\ref{fig:ladder_eps}, we plot the heat capacity for different $\varepsilon$ and fixed $\alpha = 0.5$. 
Again, the heat capacity always vanishes with temperature going to zero; again, for every epsilon, the heat capacity has a minimum or maximum in the low-temperature region.  
Just like in the case of the ring, the behavior of $C(T)$ at low $T$ is very sensitive to changes in $\varepsilon$.

For $N=2$, the analytical solution obtained from the graphical method is given in Appendix \ref{sec:app_analy}.

\section{Conclusion}
Thermal properties of active and driven systems are of great interest for probing the nature of a steady nonequilibrium condition.  Excess work and excess heat are crucial quantities here and the matrix-forest theorem gives rise to a graphical representation of them.
This graphical method actually yields an \emph{exact} calculation and no approximations need be made in finding $\rho$ or $V$. That is especially helpful for small systems, as they already generate nontrivial steady states which can become far-from-equilibrium.  The graphical method may then even produce analytical expressions for the heat capacities of such systems. More generally, it enables to study the effect of the geometry analytically in different limits. For example, in all steps of the calculation we can directly see the influence of the rates over each edge on the final result. This is to be contrasted against the more standard, yet less transparent method of calculating the spectrum of $L$, and using that to calculate the excess heat.

The proposed method is general and allows the calculation of heat capacities for any dynamics on any graph.
As an example we calculated the heat capacity for active particles in different geometries, represented as (double) graphs. 
The obtained heat capacities show  behavior not typically found in equilibrium systems, notably the frequent occurrence of minima and maxima, and negative heat capacities.  Those fascinating features are waiting to be confirmed experimentally as well.

\begin{acknowledgments}
We thank Christian Maes for many discussions and suggestions.
\end{acknowledgments}

\bibliography{graphical-hc}

\onecolumngrid
\newpage
\appendix

\section{Analytical solution for the active two-level system}\label{sec:app_analy}
In this section we give analytical results for the two-level system with a switch (the $N=2$ Sisyphus random walk).
We call the tumbling rate $\alpha$, the driving strength $\varepsilon$ and the energy spacing $\Delta$.
The transition rates between the two levels are given by \eqref{eq:bounded_rate_general}.
As indicated in section \ref{sec:alg}, the calculation of the heat capacity requires knowledge of $\rho^T(i)$ and $V^T(i)$. 
These can be calculated by employing the graphical methods laid out in section \ref{sec:alg}.

The spanning trees rooted in $i$, $\mathcal{T}_i$ for the two-level system are given by Fig.~\ref{fig:spanning_trees}, relabeling the vertices as needed for roots other than $i$.

The spanning forests $\mathcal{F}^{i\to i}$ are given in Fig.~\ref{fig:spanning_forests}.
Although the general cases for $\mathcal{F}^{i\to j}$ are not given explicitly, they are all subsets of $\mathcal{F}^{i\to i}$, and can be found by eliminating the forests not connecting $i$ and $j$.

To find an expression for the stationary distribution $\rho^T(i)$, we use \eqref{eq:graph_rep_rho},

\begin{equation}
    \begin{aligned}
    \rho^T(i)&=\frac{\mathcal{M}(\mathcal{T}_i)}{\sum_i \mathcal{M}(\mathcal{T}_i)}\\
    &=\frac{1}{2 (2 \alpha +1) \left(e^{\Delta /T}+e^{\frac{\varepsilon }{2 T}}\right) \left(e^{\frac{2 \Delta +\varepsilon }{2 T}}+1\right)}\begin{pmatrix}
    e^{\Delta /T} \left(\alpha +(2 \alpha +1) e^{\frac{2 \Delta +\varepsilon }{2 T}}+\alpha  e^{\varepsilon /T}+1\right)\\
    (\alpha +1) e^{\frac{\Delta +\varepsilon }{T}}+\alpha  e^{\Delta /T}+(2 \alpha +1) e^{\frac{\varepsilon }{2 T}}\\
    e^{\Delta /T} \left(\alpha +(2 \alpha +1) e^{\frac{2 \Delta +\varepsilon }{2 T}}+(\alpha +1) e^{\varepsilon /T}\right)\\
    \alpha  e^{\frac{\Delta +\varepsilon }{T}}+(\alpha +1) e^{\Delta /T}+(2 \alpha +1) e^{\frac{\varepsilon }{2 T}}
    \end{pmatrix}
    \end{aligned}
\end{equation}

and for $V^T(i)$, we use \eqref{eq:graph_rep_V},

\begin{equation}
\begin{aligned}
    V^T(i) &= \frac{\sum_{j} \mathcal{M}(\mathcal{F}^{i\to j}) f(j)}{\sum_j \mathcal{M}(\mathcal{T}_j)} \\ 
    &=-\frac{\varepsilon}{(2 \alpha +1)^2 \left(e^{\Delta /T}+e^{\frac{\varepsilon }{2 T}}\right) \left(e^{\frac{2 \Delta +\varepsilon }{2 T}}+1\right)}
    \begin{pmatrix}
     \left((\alpha +1) e^{\frac{\Delta +\varepsilon }{T}}+\alpha  e^{\Delta /T}+(2 \alpha +1) e^{\frac{\varepsilon }{2 T}}\right)\\
    - e^{\Delta /T} \left(\alpha +(2 \alpha +1) e^{\frac{2 \Delta +\varepsilon }{2 T}}+\alpha  e^{\varepsilon /T}+1\right)\\
    -  \left(\alpha  e^{\frac{\Delta +\varepsilon }{T}}+(\alpha +1) e^{\Delta /T}+(2 \alpha +1) e^{\frac{\varepsilon }{2 T}}\right)\\
     e^{\Delta /T} \left(\alpha +(2 \alpha +1) e^{\frac{2 \Delta +\varepsilon }{2 T}}+(\alpha +1) e^{\varepsilon /T}\right)
    \end{pmatrix}.
\end{aligned}
\end{equation}
Then to calculate the heat capacity \eqref{eq:heatcapacity},
\begin{equation}
    \begin{aligned}
        C(T) &= \frac{\dif\langle E\rangle_T}{\dif T} - \Big\langle\frac{\dif V^{T}}{\dif T}\Big\rangle_T\\
        &= \sum_i E(i)\frac{\dif \rho^T(i)}{\dif T} - \sum_i \rho^T(i)\frac{\dif V^T(i)}{\dif T}\,.
    \end{aligned}
\end{equation}
All put together, the result is
\begin{equation*}
    C(T) = \frac{2\left(\left(\Delta+2\alpha\Delta\right)^2+2\varepsilon^2\right)\left(1+\cosh\left(\frac{\Delta}{T}\right)\cosh\left(\frac{\varepsilon}{2T}\right)\right)-\left(9+4\alpha\left(1+\alpha\right)\right)\Delta\varepsilon\sinh\left(\frac{\Delta}{T}\right)\sinh\left(\frac{\varepsilon}{2T}\right)}{4T^2\left(1+2\alpha\right)^2\left(\cosh\left(\frac{3}{4T}\right)+\cosh\left(\frac{1}{T}\right)\right)^2}.
\end{equation*}

\section{Finding Forests}\label{sec:forest_algorithm}
Here we briefly sketch an algorithm which can find all sets of spanning forests $\mathcal{F}^{j\to i}$ for any given connected graph. The forests are \emph{grown} by repeatedly adding edges according to certain rules, an important fact is that any spanning forest with two trees will contain $n-2$ edges. The steps are as follows
\begin{enumerate}
    \item\label{step:dir_tree} Find all trees of size smaller than $n-2$ connecting vertex $j$ to vertex $i$ in the direction $j\to i$.
    \item\label{step:init_growth} Choose a tree from step \ref{step:dir_tree}, and choose an additional root, other than $i$.
    \item\label{step:grow} \emph{Grow} the forest from the directed tree from step \ref{step:dir_tree} and the chosen root. The growing has some sub-steps, which are
    \begin{enumerate}
        \item\label{substep:grow} Find all edges which can `legally' be added to the forest, respecting the chosen roots and the directedness of the trees.
        \item For each edge which can be added, add the edge and put all the newly found incomplete forests in a \emph{stack}, i.e., a container for unfinished forests. If the forest has $n-2$ edges, do not put it in the stack, but keep it in the set of completed forests.
        \item Take a forest from the stack and return to step \ref{substep:grow} with that forest. Repeat this procedure until the stack is empty. 
    \end{enumerate}
    \item Repeat step \ref{step:grow} for each combination of directed trees from \ref{step:dir_tree} and additional roots.
\end{enumerate}

To find all spanning \emph{trees} rooted in some $i$, $\mathcal{T}_i$, some minor modifications in the algorithm above are in order: step \ref{step:dir_tree} and \ref{step:init_growth} can be skipped, and the growth can start immediately from the root, $i$. The growth phase remains the same, bar that the number of edges in a spanning tree is $n-1$ instead of $n-2$.

The code used to generate the trees and forests for calculations in this paper is available in \cite{gitlab}. 
\end{document}

%% file: figures/double_graph.tex
\begin{tikzpicture}[auto,scale=2, centerorigin]
    \def \num {2}
    \def \numm {1}
    \def \step {0.2}
    \def \width {1.2}
    
    \foreach \x in {1,...,\num}{
        
        \node[shape=circle,fill=black, inner sep=0.05cm] (x\x0) at ({-\width/2},{\x*\step}) {};
        \node[shape=circle,fill=black, inner sep=0.05cm] (x\x1) at ({\width/2},{\x*\step}) {};
    }
    
    \foreach \x in {1,...,\numm}{
        \pgfmathtruncatemacro{\y}{\x+1}
        \draw[thick, decoration={markings, mark=at position 0.55 with {\arrow{Stealth}}}, postaction={decorate}] (x\x1)--(x\x0);
        \draw[thick, decoration={markings, mark=at position 0.55 with {\arrow{Stealth}}}, postaction={decorate}] (x\y0)--(x\y1);
        \draw[thick] (x\x0) -- (x\y0);
        \draw[thick] (x\x1) -- (x\y1);
        
    }
    
    \draw[draw=none] (x20) -- (x21) node [midway, above] {$k_{+}(x,y)$};
    \draw[draw=none] (x10) -- (x11) node [midway, below] {$k_{-}(x,y)$};
    \draw[draw=none] (x10) -- (x20) node [midway, left] {$\alpha$};
    \draw[draw=none] (x11) -- (x21) node [midway, right] {$\alpha$};
\end{tikzpicture}

%% file: figures/spanning_trees_4.tex
\begin{subfigure}{.24\linewidth}
    \begin{tikzpicture}[auto,every node/.style={draw,shape=circle,fill=black, inner sep=0.05cm},scale=2]
        \node[label= above left:{$i$}] (x) at (0,1) {};
        \node[label= above right:{$j$}] (y) at (1,1) {};
        \node[label= below right:{$k$}] (z) at (1,0) {};
        \node[label= below left:{$l$}] (u) at (0,0) {};
        
        \draw[thick, decoration={markings, mark=at position 0.55 with {\arrow{Stealth}}}, postaction={decorate}] (u) -- (x);
        \draw[thick, decoration={markings, mark=at position 0.55 with {\arrow{Stealth}}}, postaction={decorate}] (y) -- (x);
        \draw[thick, decoration={markings, mark=at position 0.55 with {\arrow{Stealth}}}, postaction={decorate}] (z) -- (y);
        \draw[color=lightgray, line width=0.10mm] (u) -- (z);
        
    \end{tikzpicture}
\end{subfigure}
\begin{subfigure}{.24\linewidth}
    \begin{tikzpicture}[auto,every node/.style={draw,shape=circle,fill=black, inner sep=0.05cm},scale=2]
        \node[] (x) at (0,1) {};
        \node[] (y) at (1,1) {};
        \node[] (z) at (1,0) {};
        \node[] (u) at (0,0) {};
        
        \draw[thick, decoration={markings, mark=at position 0.55 with {\arrow{Stealth}}}, postaction={decorate}] (y) -- (x);
        \draw[thick, decoration={markings, mark=at position 0.55 with {\arrow{Stealth}}}, postaction={decorate}] (z) -- (y);
        \draw[thick, decoration={markings, mark=at position 0.55 with {\arrow{Stealth}}}, postaction={decorate}] (u) -- (z);
        \draw[color=lightgray, line width=0.10mm] (u) -- (x);
        
    \end{tikzpicture}
\end{subfigure}
\begin{subfigure}{.24\linewidth}
    \begin{tikzpicture}[auto,every node/.style={draw,shape=circle,fill=black, inner sep=0.05cm},scale=2]
        \node[] (x) at (0,1) {};
        \node[] (y) at (1,1) {};
        \node[] (z) at (1,0) {};
        \node[] (u) at (0,0) {};
        
        \draw[thick, decoration={markings, mark=at position 0.55 with {\arrow{Stealth}}}, postaction={decorate}] (y) -- (x);
        \draw[thick, decoration={markings, mark=at position 0.55 with {\arrow{Stealth}}}, postaction={decorate}] (u) -- (x);
        \draw[thick, decoration={markings, mark=at position 0.55 with {\arrow{Stealth}}}, postaction={decorate}] (z) -- (u);
        \draw[color=lightgray, line width=0.10mm] (y) -- (z);
        
    \end{tikzpicture}
\end{subfigure}
\begin{subfigure}{.24\linewidth}
    \begin{tikzpicture}[auto,every node/.style={draw,shape=circle,fill=black, inner sep=0.05cm},scale=2]
        \node[] (x) at (0,1) {};
        \node[] (y) at (1,1) {};
        \node[] (z) at (1,0) {};
        \node[] (u) at (0,0) {};
        
        \draw[thick, decoration={markings, mark=at position 0.55 with {\arrow{Stealth}}}, postaction={decorate}] (u) -- (x);
        \draw[thick, decoration={markings, mark=at position 0.55 with {\arrow{Stealth}}}, postaction={decorate}] (y) -- (z);
        \draw[thick, decoration={markings, mark=at position 0.55 with {\arrow{Stealth}}}, postaction={decorate}] (z) -- (u);
        \draw[color=lightgray, line width=0.10mm] (x) -- (y);
        
    \end{tikzpicture}
\end{subfigure}

%% file: figures/spanning_forests_4.tex
\begin{tikzpicture}[auto,every node/.style={draw,shape=circle,fill=black, inner sep=0.05cm},scale=2]
        \def\d{1.5}
        \def\h{1.5}

        \node[label= above left:{$i$}] (x1) at (0,{1+\h}) {};
        \node[label= above right:{$j$}] (y1) at (1,{1+\h}) {};
        \node[label= below right:{$k$}] (z1) at (1,{0+\h}) {};
        \node[label= below left:{$l$}] (u1) at (0,{0+\h}) {};
        
        \node[] (x2) at ({0+\d},{1+\h}) {};
        \node[] (y2) at ({1+\d},{1+\h}) {};
        \node[] (z2) at ({1+\d},{0+\h}) {};
        \node[] (u2) at ({0+\d},{0+\h}) {};
        
        \node[] (x3) at ({0+2*\d},{1+\h}) {};
        \node[] (y3) at ({1+2*\d},{1+\h}) {};
        \node[] (z3) at ({1+2*\d},{0+\h}) {};
        \node[] (u3) at ({0+2*\d},{0+\h}) {};
        
        \node[] (x4) at ({0+3*\d},{1+\h}) {};
        \node[] (y4) at ({1+3*\d},{1+\h}) {};
        \node[] (z4) at ({1+3*\d},{0+\h}) {};
        \node[] (u4) at ({0+3*\d},{0+\h}) {};
        
        \node[] (x5) at ({0+4*\d},{1+\h}) {};
        \node[] (y5) at ({1+4*\d},{1+\h}) {};
        \node[] (z5) at ({1+4*\d},{0+\h}) {};
        \node[] (u5) at ({0+4*\d},{0+\h}) {};
        
        \node[] (x21) at (0,{1+0*\h}) {};
        \node[] (y21) at (1,{1+0*\h}) {};
        \node[] (z21) at (1,{0+0*\h}) {};
        \node[] (u21) at (0,{0+0*\h}) {};
        
        \node[] (x22) at ({0+\d},{1+0*\h}) {};
        \node[] (y22) at ({1+\d},{1+0*\h}) {};
        \node[] (z22) at ({1+\d},{0+0*\h}) {};
        \node[] (u22) at ({0+\d},{0+0*\h}) {};
        
        \node[] (x23) at ({0+2*\d},{1+0*\h}) {};
        \node[] (y23) at ({1+2*\d},{1+0*\h}) {};
        \node[] (z23) at ({1+2*\d},{0+0*\h}) {};
        \node[] (u23) at ({0+2*\d},{0+0*\h}) {};
        
        \node[] (x24) at ({0+3*\d},{1+0*\h}) {};
        \node[] (y24) at ({1+3*\d},{1+0*\h}) {};
        \node[] (z24) at ({1+3*\d},{0+0*\h}) {};
        \node[] (u24) at ({0+3*\d},{0+0*\h}) {};
        
        \node[] (x25) at ({0+4*\d},{1+0*\h}) {};
        \node[] (y25) at ({1+4*\d},{1+0*\h}) {};
        \node[] (z25) at ({1+4*\d},{0+0*\h}) {};
        \node[] (u25) at ({0+4*\d},{0+0*\h}) {};
        
        \draw[thick, decoration={markings, mark=at position 0.55 with {\arrow{Stealth}}}, postaction={decorate}] (z1) -- (u1);
        \draw[thick, decoration={markings, mark=at position 0.55 with {\arrow{Stealth}}}, postaction={decorate}] (y1) -- (x1);
        
        \draw[color=lightgray, line width=0.10mm] (u1) -- (x1);
        \draw[color=lightgray, line width=0.10mm] (z1) -- (y1);
        
        \draw[thick, decoration={markings, mark=at position 0.55 with {\arrow{Stealth}}}, postaction={decorate}] (y2) -- (z2);
        \draw[thick, decoration={markings, mark=at position 0.55 with {\arrow{Stealth}}}, postaction={decorate}] (z2) -- (u2);
        
        \draw[color=lightgray, line width=0.10mm] (u2) -- (x2);
        \draw[color=lightgray, line width=0.10mm] (x2) -- (y2);
        
        \draw[thick, decoration={markings, mark=at position 0.55 with {\arrow{Stealth}}}, postaction={decorate}] (u3) -- (x3);
        \draw[thick, decoration={markings, mark=at position 0.55 with {\arrow{Stealth}}}, postaction={decorate}] (y3) -- (z3);
        
        \draw[color=lightgray, line width=0.10mm] (y3) -- (x3);
        \draw[color=lightgray, line width=0.10mm] (u3) -- (z3);
        
        \draw[thick, decoration={markings, mark=at position 0.55 with {\arrow{Stealth}}}, postaction={decorate}] (u4) -- (x4);
        \draw[thick, decoration={markings, mark=at position 0.55 with {\arrow{Stealth}}}, postaction={decorate}] (y4) -- (x4);
        
        \draw[color=lightgray, line width=0.10mm] (y4) -- (z4);
        \draw[color=lightgray, line width=0.10mm] (u4) -- (z4);
        
        \draw[thick, decoration={markings, mark=at position 0.55 with {\arrow{Stealth}}}, postaction={decorate}] (z5) -- (u5);
        \draw[thick, decoration={markings, mark=at position 0.55 with {\arrow{Stealth}}}, postaction={decorate}] (u5) -- (x5);
        
        \draw[color=lightgray, line width=0.10mm] (y5) -- (z5);
        \draw[color=lightgray, line width=0.10mm] (x5) -- (y5);
        
        \draw[thick, decoration={markings, mark=at position 0.55 with {\arrow{Stealth}}}, postaction={decorate}] (u21) -- (z21);
        \draw[thick, decoration={markings, mark=at position 0.55 with {\arrow{Stealth}}}, postaction={decorate}] (y21) -- (x21);
        
        \draw[color=lightgray, line width=0.10mm] (y21) -- (z21);
        \draw[color=lightgray, line width=0.10mm] (x21) -- (u21);
        
        \draw[thick, decoration={markings, mark=at position 0.55 with {\arrow{Stealth}}}, postaction={decorate}] (u22) -- (z22);
        \draw[thick, decoration={markings, mark=at position 0.55 with {\arrow{Stealth}}}, postaction={decorate}] (z22) -- (y22);
        
        \draw[color=lightgray, line width=0.10mm] (y22) -- (x22);
        \draw[color=lightgray, line width=0.10mm] (x22) -- (u22);
        
        \draw[thick, decoration={markings, mark=at position 0.55 with {\arrow{Stealth}}}, postaction={decorate}] (u23) -- (x23);
        \draw[thick, decoration={markings, mark=at position 0.55 with {\arrow{Stealth}}}, postaction={decorate}] (z23) -- (y23);
        
        \draw[color=lightgray, line width=0.10mm] (y23) -- (x23);
        \draw[color=lightgray, line width=0.10mm] (z23) -- (u23);
        
        \draw[thick, decoration={markings, mark=at position 0.55 with {\arrow{Stealth}}}, postaction={decorate}] (u24) -- (z24);
        \draw[thick, decoration={markings, mark=at position 0.55 with {\arrow{Stealth}}}, postaction={decorate}] (y24) -- (z24);
        
        \draw[color=lightgray, line width=0.10mm] (y24) -- (x24);
        \draw[color=lightgray, line width=0.10mm] (x24) -- (u24);
        
        \draw[thick, decoration={markings, mark=at position 0.55 with {\arrow{Stealth}}}, postaction={decorate}] (y25) -- (x25);
        \draw[thick, decoration={markings, mark=at position 0.55 with {\arrow{Stealth}}}, postaction={decorate}] (z25) -- (y25);
        
        \draw[color=lightgray, line width=0.10mm] (u25) -- (x25);
        \draw[color=lightgray, line width=0.10mm] (z25) -- (u25);
        
\end{tikzpicture}

%% file: figures/dring.tex
\begin{tikzpicture}[auto,scale=1]
    \def \num {5}
    \def \numm {4}
    \def \ang {360/\num}
    \def \offset {18}
    \def \inner {1.2}
    \def \outer {1.8}
    
    \foreach \x in {1,...,\num}{
        \node[shape=circle,fill=black, inner sep=0.05cm] (x\x0) at ({\x*\ang+\offset}:\inner) {};
        \node[shape=circle,fill=black, inner sep=0.05cm] (x\x1) at ({\x*\ang+\offset}:\outer) {};
        \draw[thick] (x\x0)--(x\x1);
    }
    
    \foreach \x in {1,...,\numm}{
        \pgfmathtruncatemacro{\y}{\x+1}
        \draw[thick, decoration={markings, mark=at position 0.6 with {\arrow{Stealth}}}, postaction={decorate}] (x\x0)--(x\y0);
        \draw[thick, decoration={markings, mark=at position 0.6 with {\arrow{Stealth}}}, postaction={decorate}] (x\y1)--(x\x1);
        
    }
    
    \draw[thick, decoration={markings, mark=at position 0.6 with {\arrow{Stealth}}}, postaction={decorate}] (x50)--(x10);
    \draw[thick, decoration={markings, mark=at position 0.6 with {\arrow{Stealth}}}, postaction={decorate}] (x11)--(x51);
    
    \draw[draw=none] (x20) -- (x21) node [midway, above right] {$\alpha$};
    \draw[draw=none] (x10) -- (x20) node [midway, below right] {$\varepsilon$};
    \draw[draw=none] (x11) -- (x21) node [midway, above left] {$\varepsilon$};
\end{tikzpicture}

%% file: figures/ladder.tex
\begin{tikzpicture}[auto,centerorigin,scale=1]
    \def \num {4}
    \def \numm {3}
    \def \step {1}
    \def \width {1.2}
    
    \foreach \x in {1,...,\num}{
        
        \if \x 1{
            \draw[thin,dashed,color=gray] (0,{\x*\step}) -- ({\width*0.7},{\x*\step}) node [right,color=black] {$E_0$};
                }
        \else{
            \draw[thin,dashed,color=gray] (0,{\x*\step}) -- ({\width*0.7},{\x*\step}) node [right,color=black] {${\x}E_0$};
            }
        \fi
        
        \node[shape=circle,fill=black, inner sep=0.05cm] (x\x0) at (-\width,{\x*\step}) {};
        \node[shape=circle,fill=black, inner sep=0.05cm] (x\x1) at (0,{\x*\step}) {};
        \draw[thick] (x\x0)--(x\x1);
    }
    
    \foreach \x in {1,...,\numm}{
        \pgfmathtruncatemacro{\y}{\x+1}
        \draw[thick, decoration={markings, mark=at position 0.65 with {\arrow{Stealth}}}, postaction={decorate}] (x\x0)--(x\y0);
        \draw[thick, decoration={markings, mark=at position 0.65 with {\arrow{Stealth}}}, postaction={decorate}] (x\y1)--(x\x1);
        
    }
    
    \draw[draw=none] (x20) -- (x21) node [midway, above] {$\alpha$};
    \draw[draw=none] (x10) -- (x20) node [midway, right] {$\varepsilon$};
    \draw[draw=none] (x11) -- (x21) node [midway, left] {$\varepsilon$};
\end{tikzpicture}